\documentclass[preprint,showpacs,preprintnumbers,amsmath,amssymb,nofootinbib]{revtex4}

\usepackage{graphicx}% Include figure files
\usepackage{dcolumn}% Align table columns on decimal point
\usepackage{bm}% bold math

\def\cO#1{{\cal{O}}\left(#1\right)}

\begin{document}

\preprint{FERMILAB-PUB-05-238-T}

\title{Virtual QCD corrections to Higgs boson \\plus four parton processes}

\author{R.K. Ellis}
\email{ellis@fnal.gov}
\author{W.T. Giele}%
\email{giele@fnal.gov}

\author{G. Zanderighi}
\email{zanderi@fnal.gov}
\affiliation{
Fermilab\\
Batavia, IL 60510, USA}

\begin{abstract}
  We report on the calculation of virtual processes contributing to
  the production of a Higgs boson and two jets in hadron-hadron
  collisions.  The coupling of the Higgs boson to gluons, via a
  virtual loop of top quarks, is treated using an effective theory,
  valid in the large top quark mass limit. The calculation is
  performed by evaluating one-loop diagrams in the effective theory.
  The primary method of calculation is a numerical evaluation of the
  virtual amplitudes as a Laurent series in $D-4$, where $D$ is the
  dimensionality of space-time.  For the cases $H \rightarrow
  q\bar{q}q\bar{q}$ and $H \rightarrow q\bar{q}q'\bar{q}'$ we confirm
  the numerical results by an explicit analytic calculation.
\end{abstract}

\pacs{12.38.Bx, 14.80.Bn}
\maketitle

\newcommand{\beq}{\begin{equation}}
\newcommand{\eeq}{\end{equation}}
\newcommand{\bea}{\begin{eqnarray}}
\newcommand{\eea}{\end{eqnarray}}
\newcommand{\nn}{\nonumber}

\def\li{{\rm Li_2}}
\def\Ll{{\rm L}}
\def\Ls{\mathop{\rm Ls}\nolimits}
\def\Lsnew{{\widetilde{\rm Ls}}}
\def\Cg{c_\Gamma}
\def\MSbar{\overline{\mbox{\small MS}}}
\def\ku{k_1}
\def\kd{k_2}
\def\kt{k_3}
\def\kq{k_4}
\def\sud{s_{12}}
\def\sut{s_{13}}
\def\suq{s_{14}}
\def\sdt{s_{23}}
\def\sdq{s_{24}}
\def\stq{s_{34}}
\def\sudt{s_{123}}
\def\sudq{s_{124}}
\def\sutq{s_{134}}
\def\sdtq{s_{234}}
\def\mhsq{M_H^2}
\def\Nc{N_c}
\def\eqn#1{Eq.~(\ref{#1})}
\def\eqns#1#2{Eqs.~(\ref{#1}) and~(\ref{#2})}
\def\eqnss#1#2{Eqs.~(\ref{#1})-(\ref{#2})}
\def\fig#1{Fig.~{\ref{#1}}}
\def\sec#1{Section~{\ref{#1}}}
\def\app#1{Appendix~\ref{#1}}
\def\tab#1{Table~\ref{#1}}

%%%%%%%%%%%%%%%%%%%%%%%%%%%%%%%%%%%%%%%%%%%%%%%%%%%%%%%%%%%%%

\newcommand{\ttbs}{\char'134}           % \backslash for \tt (Nucl.Phys. :)%
\newcommand\fverb{\setbox\pippobox=\hbox\bgroup\verb}
\newcommand\fverbdo{\egroup\medskip\noindent%
                        \fbox{\unhbox\pippobox}\ }
\newcommand\fverbit{\egroup\item[\fbox{\unhbox\pippobox}]}
\newbox\pippobox
%   ...                                                                    %
%%%%%%%%%%%%%%%%%%%%%%%%%%%%%%%%%%%%%%%%%%%%%%%%%%%%%%%%%%%%%%%%%%%%%%%%%%%%

\def\spa#1.#2{\left\langle#1\,#2\right\rangle}
\def\spb#1.#2{\left[#1\,#2\right]}
\def    \sapp#1#2#3#4{{\langle #1 | (#2+#3) |#4  \rangle} }

\def\feynsl#1{
  \setbox0=\hbox{/} \setbox1=\hbox{$#1$}
  \dimen0=\wd0 \advance\dimen0 by -\wd1 \divide\dimen0 by 2
  \ifdim\wd0>\wd1 \raise.15ex\copy0\kern-\wd0\kern\dimen0\copy1\kern\dimen0
  \else \kern-\dimen0\raise.15ex\copy0\kern-\dimen0\kern-\wd1\copy1\fi}

\def\dprod#1#2{{\left({#1}\!\cdot\!{#2}\right)}}
\def\iscol#1#2{{\buildrel#1\parallel#2\over\longrightarrow}}

%%%%%%%%%%%%%%%%%%%%%%%%%%%%%%%%%%%%%%%%%%%%%%%%%%%%%%%%%%%%%

\def\tr{\mathop{\rm tr}\nolimits}
\def\top{{\rm top}}
\def\mt{M_\top}
\def\eff{{\rm eff}}
\def\gev{{\rm GeV}}
\def\ord{{\cal O} }
\def\cM{{\cal M}}
\def\cL{{\cal L}}
\def\la{\langle}
\def\ra{\rangle}
\def\sep{\mbox{$\,|\;$}}
\def\ib{{\bar\imath}}
\def\jb{{\bar\jmath}}
\def\qb{{\bar q}}
\def\Qb{{\overline Q}}

\newcommand\sss{\scriptscriptstyle}
\newcommand\as{\alpha_{\sss S}} 
\newcommand\gs{g_{\sss S}}
\def\CF{C_{\sss F}}
\newcommand\ph{p_{\sss H}}
\newcommand\pt{p^{\sss\rm T}}
\newcommand\ptj{p^{\sss\rm T}_j}
\newcommand\ptmin{p^{\sss\rm T}_{min}}
\newcommand\mh{M_{\sss H}}
\newcommand\yh{y_{\sss H}}
\newcommand\yrel{y_{\rm rel}}
\newcommand\sh{s_{j\sss H}}
\newcommand\sah{s_{j_1\sss H}}
\newcommand\sbh{s_{j_2\sss H}}
\def\e{\epsilon}

%%%%%%%%%%%%%%%%%%%%%%%%%%%%%%%%%%%%%%%%%%%%%%%%%%%%%%%%%%%%%
% macros for the spinor products

\def    \br(#1,#2)          {\mbox{$\langle #1 \, #2 \rangle$}}
% use: write \br(1,3) .....
\def    \sq(#1,#2)          {\mbox{$\left[  #1 \, #2 \right]$}}

%%%%%%%%%%%%%%%%%%%%%%%%%%%%%%%%%%%%%%%%%%%%%%%%%%%%%%%%%%%%%

\section{Introduction}

\begin{figure}
\includegraphics[angle=270,scale=0.8]{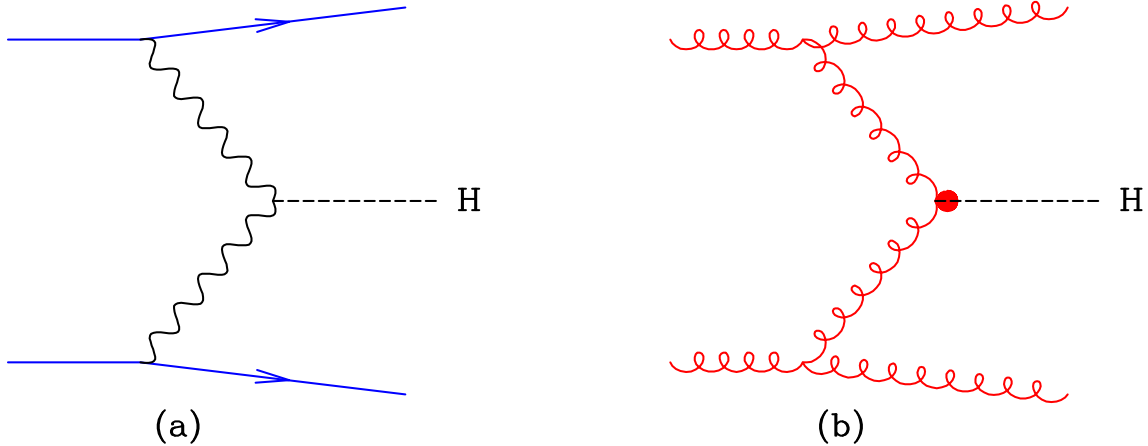}

\caption{\label{fusion}(a) Lowest order process for vector boson fusion.
(b) Example of a diagram contributing to the gluon fusion process 
in association with two jets.}
\end{figure}

In this paper we study the production of a standard model Higgs boson
in association with two jets.  This is one of the most promising
discovery channels at the LHC especially for a Higgs boson with a
mass 
in the range $110$\,GeV$<M_H<180$\,GeV.  At Born level there are two classes of
processes which contribute, as illustrated in Fig.~\ref{fusion}.

In Fig.~\ref{fusion}(a), the Higgs is produced via vector boson
fusion, while in Fig.~\ref{fusion}(b) the coupling of the Higgs boson
to gluons is mediated by a top quark loop. In the limit in which the
mass of the top quark tends to infinity the coupling can be treated
using an effective theory as described below.  We shall refer to this
process as the gluon fusion process. Notice that the external gluons
lines in Fig.~\ref{fusion}(b) could as well be replaced by quarks.

The final aim of this study is the calculation of the Higgs + $2$-jet rate,
at next-to-leading (NLO) order, where the Higgs is produced using 
the effective coupling to gluons, 
\beq 
\cL_\eff = \frac{1}{4} A (1+\Delta) H G^a_{\mu \nu} G^{a\,\mu \nu}\, .
\label{eq:Leff}
\eeq
In Eq.~(\ref{eq:Leff}), $G^a_{\mu \nu}$ is the field strength of the
gluon field and $H$ is the Higgs-boson field.
The effective coupling $A$ is given by 
\beq
A = \frac{g^2}{12 \pi^2 v}\,,
\eeq
where $g$ is the bare strong coupling and $v$ is the vacuum expectation
value parameter, $v^2=(G_F\sqrt{2})^{-1}=(246~\gev)^2$. 
The finite $O(g^2$) correction to the effective operator
has been calculated~\cite{Djouadi:1991tk,Dawson:1990zj}
\beq \label{Delta}
\Delta = \frac{11 g^2}{16 \pi^2}\,.
\eeq
The full NLO result will require the evaluation of the virtual
corrections to the Higgs + 4 parton processes, which are the subject
of this paper, the calculation of the tree graph rates from the Higgs
+ 5 partons amplitudes already given in
refs.~\cite{DelDuca:2004wt,Badger:2004ty,Dixon:2004za} and the
calculation of a set of subtraction terms.  The subtractions remove
singularities present in the real emission diagrams in the regions of
soft and collinear emission.  After integration over the momentum of
the un observed parton they are added back to the virtual emission
diagrams and cancel the singularities in those virtual terms.

We believe this calculation would be a useful addition to the
literature for several reasons. First, the effective Lagrangian
approach appears to be valid for light Higgs boson
mass if the transverse momentum of the associated
jets is less than the top quark mass~\cite{DelDuca:2001eu,DelDuca:2001fn}.  
Second, this process
constitutes a `background' to the experimentally interesting vector
boson fusion process, Fig~\ref{fusion}(a).  A complete NLO calculation
will improve knowledge of this `background' process.  In addition,
because the vector boson fusion process has a well determined
normalization, it is one of the most accurate sources of information
about the couplings of the Higgs boson at the
LHC~\cite{Zeppenfeld:2000td}.  An uncontrolled background from gluon
fusion process could compromise that measurement.  For a comprehensive
review of standard model Higgs physics, see
ref.~\cite{Djouadi:2005gi}.

Note that the process calculated in this paper is distinguished from
the vector boson process, Fig~\ref{fusion}(a), by the presence of
colored particles exchanged in the $t$-channel.  The exchange of color
charge generates extra jet activity in the central region, allowing
discrimination against this process by a jet veto. Although the
efficacy of such a veto will finally have to be determined by
experiment, it will still be interesting to see how this works at the
parton level with a full NLO calculation~\footnote{To a limited extent
  this has been looked at in ref.~\cite{DelDuca:2004wt}.  However in a
  tree graph calculation one cannot look at the effects of finite jet
  size or of the central jet veto.}.

In the large top quark mass limit, virtual corrections have been
considered in the effective theory by previous authors. Loop
corrections to the process $H \rightarrow gg $ are considered at
one-loop level in ref.~\cite{Dawson:1990zj} and at two loop level in
refs.~\cite{Harlander:2000mg,Anastasiou:2002yz}.  The results for the
processes $H \rightarrow ggg $ and $H \rightarrow q \bar{q} g $ are
given in refs.~\cite{Schmidt:1997wr,Ravindran:2002dc}. In the
following we shall describe results for the virtual corrections to the
processes
\begin{eqnarray} \label{fourprocesses1} 
A)\;\; H &\rightarrow& q \bar{q} q' \bar{q}' \; ,\\ 
\label{fourprocesses2} 
B)\;\; H &\rightarrow& q \bar{q} q \bar{q} \; ,\\ 
\label{fourprocesses3} 
C)\;\; H &\rightarrow& q \bar{q} gg  \; , \\ 
\label{fourprocesses4} 
D)\;\; H &\rightarrow& gggg \; ,
\end{eqnarray}
using the effective theory, Eq.~(\ref{eq:Leff}).

\section{Lowest order process}

\subsection{$H \rightarrow  q\bar{q}q'\bar{q}'$}
\begin{figure}
\includegraphics[angle=270,scale=0.8]{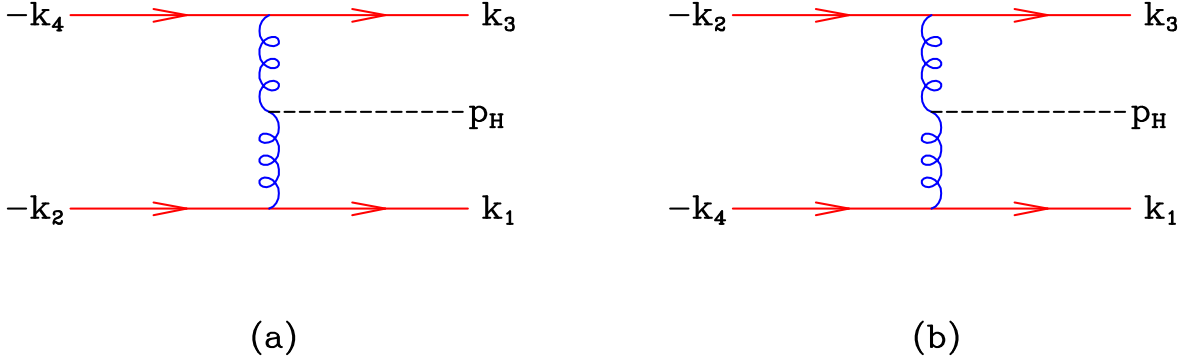}
\caption{\label{diag}(a) Lowest order process for $H \to q \bar{q} q' \bar{q}'$.
(b) Second diagram for identical quark process
$H \to q \bar{q} q \bar{q}$.}

\end{figure}

We first perform the calculation of the matrix element for the process
involving two distinct flavors of massless quarks, $q$ and $q'$,
process $A$,
\beq \label{Hqqrr}
H \rightarrow  q(\ku) + \bar{q}(\kd) + q'(\kt)  +\bar{q}'(\kq) \; .
\eeq
At Born level, only the diagram in Fig.~\ref{diag}(a) contributes. 
The color
expansion of the amplitude can be written as
\beq
M^A_0(\ku,\kd,\kt,\kq)
= \Big[\delta^{i_1}_{i_4}\delta^{i_3}_{i_2}
-\frac{1}{N_c} \delta^{i_1}_{i_2}\delta^{i_3}_{i_4}\Big]\; 
a^{(0)}(1,2,3,4)\,,
\eeq
where $i_j$ denotes the color index of the $j$th quark
and we have introduced the notation
\beq
a^{(0)}(1,2,3,4) \equiv a^{(0)}(\ku,h_1;\kd,h_2;\kt,h_3;\kq,h_4)\,, 
\eeq
where $k_i$ and $h_i$ denote the momentum and the helicity of quark
$i$.  The result for the squared matrix element summed over the spins
and colors of the final state quarks and antiquarks is then
\begin{eqnarray}
A_0(\ku,\kd,\kt,\kq) &\equiv& \sum |M_0^A (\ku,\kd,\kt,\kq)|^2 \nonumber \\
  &=&  g^4 A^2 V\Bigg[
 \frac{(\sut \sdq - \sdt \suq)^2+\sud^2 \stq^2} {\stq^2 \sud^2} 
    +\frac{(\sut - \sdq)^2+(\suq - \sdt)^2}{2 \stq \sud}\Bigg] \; .
\nonumber \\
\end{eqnarray}
The number of colors, $\Nc$, enters as $V=\Nc^2-1$, so, for the case
of SU(3), we have that $V=8$. The Lorentz invariants are defined as
$s_{ij}\equiv (k_i+k_j)^2 = 2 k_i \cdot k_j $. The momentum of the
Higgs can be eliminated in terms of the four massless momenta,
$p_H=-\ku-\kd-\kt-\kq$, so that
\beq
M_H^2= \sud+\sut+\suq+\sdt+\sdq+\stq \,.
\eeq

\subsection{$H \rightarrow  q\bar{q}q\bar{q}$}
In the case of massless quarks of identical flavor, process $B$, 
\beq \label{Hqqqq}
H \rightarrow  q(\ku) + \bar{q}(\kd) + q(\kt)  +\bar{q}(\kq) \; ,
\eeq
the Born amplitude squared is determined by the two diagrams, shown in
Fig.~\ref{diag}(a) and (b), which differ by the exchange of the final
state anti-quarks.  The color expansion of the amplitude can be
written as
\begin{eqnarray}
M^B_0(\ku,\kd,\kt,\kq)
&=& 
\Big[\delta^{i_1}_{i_4}\delta^{i_3}_{i_2}
-\frac{1}{N_c} \delta^{i_1}_{i_2}\delta^{i_3}_{i_4}\Big]\; a^{(0)}(1,2,3,4)
-\Big[\delta^{i_1}_{i_2}\delta^{i_3}_{i_4}
-\frac{1}{N_c} \delta^{i_1}_{i_4}\delta^{i_3}_{i_2}\Big]\; 
a^{(0)}(1,4,3,2)\nonumber\\
&=& M_0^A(\ku,\kd,\kt,\kq) - M_0^A(\ku,\kq,\kt,\kd)\,.
\end{eqnarray}
The result for the matrix element squared, summed over the spins 
and colors of the final state quarks and antiquarks is given by 
\begin{eqnarray}
B_0(\ku,\kd,\kt,\kq) &\equiv&
\sum |M_0^B (\ku,\kd,\kt,\kq)|^2 
\nonumber \\
&=& 
 A_0(\ku,\kd,\kt,\kq)+A_0(\ku,\kq,\kt,\kd)
+B_0^\prime(\ku,\kd,\kt,\kq)\; ,
\end{eqnarray}
where the interference term is defined as
\begin{equation}
B_0^\prime(\ku,\kd,\kt,\kq)  \equiv 
-2 \; \sum \; {\rm Re} \; \Big[M_0^A(\ku,\kd,\kt,\kq)^{\star} \; 
M_0^A(\ku,\kq,\kt,\kd) \Big] \; .
\end{equation}
The result for the lowest order interference term, $B_0^\prime$, 
is given by,
\begin{eqnarray}
B_0^\prime(\ku,\kd,\kt,\kq) &=&
    g^4 A^2 C_f \Bigg\{\Big[(\sut-\sdq)^2 (\sud \stq+\suq \sdt-\sut \sdq) 
\nonumber \\
       &-& 2 (\sut \sdq+\suq \sdt-\sud \stq) 
             (\sud \stq+\sut \sdq-\suq \sdt)\Big]
       \Bigg\}\nonumber \\
  & \times & \frac{1}{\sud \suq \sdt \stq}\,,
\end{eqnarray}
with $C_f=(\Nc^2-1)/(2 \Nc) =4/3$. 
\relax
\subsection{$H \rightarrow  q\bar{q}gg$}
We now turn to process $C$,
\beq
  H \rightarrow q(k_1)+ \bar{q}(k_2) + g(k_3) +g (k_4) \; .
\eeq
At lowest order the amplitude is given by,
\begin{eqnarray}
M_0^C &=& (T^{a_3} T^{a_4})_{i_1 i_2} c_{1}^{(0)}(1,2,3,4)+
          (T^{a_4} T^{a_3})_{i_1 i_2} c_{2}^{(0)}(1,2,3,4) \; ,
\end{eqnarray}
where $a_3,a_4$ are the color indices of the gluons and $i_1,i_2$ are 
the color indices of the quarks.
As before we have introduced notation of the form
\beq
c^{(0)}_i(1,2,3,4)\equiv 
c^{(0)}_i(\ku,h_1;\kd,h_2;\kt,\varepsilon_3;\kq,\varepsilon_4)\; , 
\eeq
where $\varepsilon_i$ is the polarization vector of gluon $i$ and 
$c^{(0)}_2(1,2,3,4)= c^{(0)}_1(1,2,4,3)$. 
Explicit forms for the three independent helicity amplitudes can be 
found, for example 
in refs.~\cite{Kauffman:1997ix,DelDuca:2004wt}. The former reference also
contains explicit results for the amplitude squared.

\subsection{$H \rightarrow  gggg$}
Lastly we consider the matrix element for the process $D$,
\beq \label{Hgggg}
H \rightarrow g(k_1)+g(k_2)+g(k_3)+g(k_4)\,.
\eeq

At lowest order the four gluon matrix element has the structure
\begin{eqnarray}
M_0^D &=& 
\sum_{\sigma \in S_4/Z_4} 
\mbox{tr}(T^{a_{\sigma(1)}} T^{a_{\sigma(2)}} 
T^{a_{\sigma(3)}} T^{a_{\sigma(4)}}) \; 
d_{1}^{(0)}(\sigma(1),\sigma(2),\sigma(3),\sigma(4))\, ,
\end{eqnarray}
where the sum runs over the six non-cyclic permutations and we have
introduced the notation
\beq
d_{i}(1,2,3,4) \equiv 
d_{i}( \ku,\varepsilon_1;\kd,\varepsilon_2;\kt,\varepsilon_3;\kq,\varepsilon_4) \; .
\eeq
The partial amplitudes satisfy the 
relations~\cite{Mangano:1987xk,Berends:1988me}
\begin{eqnarray} 
\label{eq:cycle}
d_{i}^{(0)}(1,2,3,4)= d_{i}^{(0)}(4,1,2,3)&\qquad & \mbox{cyclicity}\; ,\\
\label{eq:reflect}
d_{i}^{(0)}(1,2,3,4)= d_{i}^{(0)}(4,3,2,1)&\qquad & \mbox{reflection}\; ,\\
\label{eq:dual}
d_{i}^{(0)}(1,2,3,4)+ d_{i}^{(0)}(2,1,3,4)+d_{i}^{(0)}(2,3,1,4) =0&\qquad & 
\mbox{dual Ward identity}\,,
\end{eqnarray}
so that at Born level for fixed helicities there are only two
independent amplitudes.  Explicit expressions for the helicity
amplitudes can be found for example in
refs.~\cite{Kauffman:1997ix,DelDuca:2004wt}. The former reference also
contains explicit results for the amplitude squared.
Eqs.~(\ref{eq:cycle}) and (\ref{eq:reflect}) continue to be valid
beyond leading order~\cite{Bern:1990ux}.

\section{Higher order processes}
In order to control the divergences which will occur at higher order
we will continue the dimensionality of space-time, $D=4-2 \epsilon$.
Within the context of dimensional regularization there remain choices
of the dimensionality of internal and external gluons which are needed
to completely specify the scheme.  The most commonly adopted choices
are the conventional dimensional regularization, (CDR), the 't
Hooft-Veltman scheme, (HV)~\cite{'tHooft:1972fi}, and the
four-dimensional helicity scheme,
(FDH)~\cite{Bern:1991aq,Bern:2002zk}.
In the CDR scheme one uniformly continues all momenta and polarization
vectors to $D$ dimensions. The HV scheme differs in the treatment of
the external states, which remain four-dimensional. Finally in the FDH
scheme all states are four-dimensional, and only the internal loop
momenta are continued to $D$ dimensions.

Since we are interested in numerical evaluation, it is preferable to
consider the external quarks and gluons in four dimensions, with two
physical helicity states.  We choose to work in the 't Hooft-Veltman
scheme.  The relationship of the CDR, HV and FDH regularization
schemes has been presented in
refs.~\cite{Kunszt:1993sd,Catani:1996pk}.  It is therefore
straightforward to translate our results to another scheme.  The
details of the translation between the HV and FDH schemes are provided
in Section~\ref{numer}.

\subsection{Distinct quarks}
At next-to-leading order in the perturbative expansion, 
30 virtual diagrams contribute to the amplitude given 
in Eq.~(\ref{Hqqrr}). At one-loop level the amplitude can be decomposed 
into two independent color structures, 
\beq
M^A_1(\ku,\kd,\kt,\kq)
= \Big[\delta^{i_1}_{i_4}\delta^{i_3}_{i_2}
-\frac{1}{N_c} \delta^{i_1}_{i_2}\delta^{i_3}_{i_4}\Big]\; 
a_1^{(1)}(1,2,3,4)
+\delta^{i_1}_{i_2}\delta^{i_3}_{i_4} \; a_2^{(1)}(1,2,3,4)
\,.
\eeq
The color sub-amplitude $ a_2^{(1)}$ does not contribute at 
next-to-leading order because the interference with the color structure
of the Born amplitude vanishes.

Before renormalization we find for the
squared matrix element, summed over spin and colors of the final state
\begin{eqnarray}
\label{eq:A1}
A_1(\ku,\kd,\kt,\kq) &\equiv & 
\sum \Big(|M_0^A+M_1^A|^2 - |M_1^A|^2\Big) \nonumber \\
&=&   A_0(\ku,\kd,\kt,\kq)
 \Bigg[1
 +\frac{g^2 }{8 \pi^2} Y^A(\ku,\kd,\kt,\kq) \Bigg] 
 \nonumber \\
   &+& A^2 \frac{V}{2} 
 \frac{g^6}{8 \pi^2} 
   \Big[ X^A(\ku,\kd,\kt,\kq) + X^A(\kt,\kq,\ku,\kd) \nonumber \\
   &+&   X^A(\kd,\ku,\kq,\kt) + X^A(\kq,\kt,\kd,\ku) \Big] 
+ O(\epsilon) \; .
\end{eqnarray}
All ultraviolet and infrared singularities are in the functions
$Y(\ku,\kd,\kt,\kq)$ given by
\begin{eqnarray}
&&Y^A(\ku,\kd,\kt,\kq)=
 -\Nc  \frac{\Cg \mu^{2 \epsilon}}{\epsilon^2} 
 \Big[(-\suq)^{-\e} + (-\sdt)^{-\e} \Big] \nonumber \\
 &+& \frac{1}{\Nc}  \frac{\Cg\mu^{2 \epsilon}}{\epsilon^2} 
 \Big[(-\sud)^{-\e} + (-\stq)^{-\e} 
 -2 (-\sut)^{-\e} +2 (-\suq)^{-\e} +2 (-\sdt)^{-\e} -2 (-\sdq)^{-\e} ) \Big] 
 \nonumber \\
  &-& \frac{\Cg\mu^{2 \epsilon}}{\e} 
 \Big[3 C_f-b_0 \Big] 
  \Big[(-\sud)^{-\e}+(-\stq)^{-\e} \Big]
  - \frac{20}{9} n_f+\frac{152}{9} \Nc-16 C_f  \nonumber \\
  &+&\frac{1}{\Nc} 
  \Big[\Ls^{2{\rm m}e}_{-1}(\sutq,\sdtq;\stq,\mhsq)
     + \Ls^{2{\rm m}e}_{-1}(\sudt,\sudq;\sud,\mhsq) \Big]\nonumber \\
  &-&\frac{2}{\Nc} 
  \Big[ \Ls^{2{\rm m}e}_{-1}(\sudt,\sutq;\sut,\mhsq)
     +  \Ls^{2{\rm m}e}_{-1}(\sudq,\sdtq;\sdq,\mhsq) \Big]\nonumber \\
  &-&(\Nc-\frac{2}{\Nc} ) \Big[ \Ls^{2{\rm m}e}_{-1}(\sudq,\sutq;\suq,\mhsq)
                              + \Ls^{2{\rm m}e}_{-1}(\sudt,\sdtq;\sdt,\mhsq)
  \Big]\,,
\end{eqnarray}
where
\beq
\Cg \equiv ( 4 \pi)^\e \frac{\Gamma(1+\e)\Gamma^2(1-\e)}{\Gamma(1-2 \e)}
 = \frac{( 4 \pi)^\e}{\Gamma(1-\e)}+O(\e^3)\,,
\eeq
and 
\beq \label{b0}
b_0= \Big(\frac{11 \Nc}{3}-\frac{2 n_f}{3}\Big) \; .
\eeq
As usual $n_f$ is the number of light flavors and $\mu$ is the scale 
introduced to keep the coupling constant dimensionless in $D$ dimensions.

The finite function $X^A(\ku,\kd,\kt,\kq)$ is given by
\begin{eqnarray}
\label{eq:XA}
X^A(\ku,\kd,\kt,\kq) &=&
    \Ls_{-1}(\sud,\sut;\sudt) \frac{2}{\Nc} \; f_{1}(\kd,\ku,\kt,\kq) \nonumber \\
  &+& \Ls_{-1}(\sud,\sdt;\sudt) (\Nc-\frac{2}{\Nc}) \; f_{1}(\ku,\kd,\kt,\kq) \nonumber \\
  &+& (\frac{1}{\Nc}+\Nc) \Ll_1(\frac{-\sudt}{-\sud})  \; f_{2}(\ku,\kd,\kt,\kq) 
  +\Nc  \Ll_0(\frac{-\sudt}{-\sud}) \; f_{3}(\ku,\kd,\kt,\kq) \nonumber \\
  &+&\frac{1}{\Nc}   \Ll_0(\frac{-\sudt}{-\sud})\; f_{4}(\ku,\kd,\kt,\kq) 
  +(-\frac{\Nc}{2}+\frac{1}{\Nc})  \Ll_0({\frac{-\sudq}{-\suq}}) \; f_{5}(\ku,\kd,\kt,\kq) \nonumber \\
  &-&\frac{1}{\Nc}   \Ll_0(\frac{-\sudt}{-\sut}) \; f_{5}(\ku,\kd,\kq,\kt) 
  +\Nc \ln(\frac{-\sudt}{-\sud}) \; f_{6}(\ku,\kd,\kt,\kq) \nonumber \\
  &+&\Nc \ln(\frac{-\sudt}{-\sdt}) \; f_{7}(\ku,\kd,\kt,\kq) 
  +\Nc \ln(\frac{-\sud}{-\suq}) \; f_{8}(\ku,\kd,\kt,\kq) \nonumber \\
  &+&\frac{1}{\Nc} \ln(\frac{-\sudt}{-\sud}) \; f_{9}(\ku,\kd,\kt,\kq) 
  +\frac{1}{\Nc} \ln(\frac{-\sudt}{-\sut}) \; f_{10}(\ku,\kd,\kt,\kq) \nonumber \\
  &-&\frac{1}{\Nc} \ln(\frac{-\sudt}{-\sdt}) \; f_{10}(\kd,\ku,\kt,\kq)
  +(\Nc+\frac{1}{\Nc}) \; f_{12}(\ku,\kd,\kt,\kq)\,.   \nonumber \\
  &+&{\frac{1}{2 \Nc}
 \Big(\ln(\frac{-\sud}{-\sut})+\ln(\frac{-\sud}{-\suq})\Big)
 \Big(f_{11}(\ku,\kd,\kt,\kq)-f_{11}(\kd,\ku,\kt,\kq)\Big)}
\,.  \nonumber \\
\end{eqnarray}
The special functions coming from the loop integrals, 
$L_0,L_1, \Ls_{-1}$ and $\Ls^{2{\rm m}e}_{-1}$
are given in Appendix~\ref{functionsA}.
The explicit expression for the kinematic functions $f_i$ are given in
Appendix~\ref{functionsD}. We note that the line-reversal symmetry 
($1 \leftrightarrow 2$ and
$3 \leftrightarrow 4$) and the renaming property ($1
\leftrightarrow 3$ and $2 \leftrightarrow 4$) are manifest in
Eq.~\eqref{eq:A1}.

The UV divergences are removed in the $\MSbar$-scheme by
adding a counterterm $A_{\rm ct}$ given by 
\beq \label{counterterm1}
 A_{\rm ct}(\ku,\kd,\kt,\kq) = 
-2 \frac{\Cg}{\epsilon} b_0  \frac{g^2}{16 \pi^2} A_0(\ku,\kd,\kt,\kq)\, .
\eeq 
Additionally, there is a finite contribution, $A_{\rm fin}$, 
coming from the effective Lagrangian, Eq.~(\ref{eq:Leff}), which is
\begin{equation} \label{counterterm2}
 A_{\rm fin}(\ku,\kd,\kt,\kq) = {2} \Delta \; A_0(\ku,\kd,\kt,\kq)\,,
\end{equation} 
where $\Delta$ is given in Eq.~(\ref{Delta}).
\subsection{Identical quarks}
In the case of identical quarks, 60 diagrams contribute the
next-to-leading order process, Eq.~(\ref{Hqqqq}). 
Before renormalization we find for the squared amplitude, summed over
colors and spins, 
\begin{eqnarray}
\label{eq:B1}
B_1(\ku,\kd,\kt,\kq) &\equiv&
\sum\Big(|M_0^B+M_1^B|^2 - |M_1^B|^2\Big)\\
&=& A_1(\ku,\kd,\kt,\kq)+A_1(\ku,\kq,\kt,\kd) +B_1^\prime(\ku,\kd,\kt,\kq)\,, 
\end{eqnarray}
with $A_1$ given in~\eqref{eq:A1}. 
The result for the interference term can be written as,
\begin{eqnarray}
B_1^\prime(\ku,\kd,\kt,\kq) &=&  B_0^\prime(\ku,\kd,\kt,\kq)\Bigg[1
 +\frac{g^2 }{8 \pi^2} Y^B(\ku,\kd,\kt,\kq) 
\Bigg] \nonumber \\
&+& A^2 V \frac{g^6}{8 \pi^2} \Bigg[
  X^B(\ku,\kd,\kt,\kq) 
+ X^B(\kt,\kd,\ku,\kq)
+ X^B(\ku,\kq,\kt,\kd) 
\nonumber \\
&+& X^B(\kt,\kq,\ku,\kd)
+ X^B(\kq,\kt,\kd,\ku)
+ X^B(\kd,\kt,\kq,\ku)
\nonumber \\
&+& X^B(\kq,\ku,\kd,\kt)
+ X^B(\kd,\ku,\kq,\kt) \Bigg] + O(\epsilon) \; ,
\end{eqnarray}
where the function $Y^B$ contains all divergent terms
\begin{eqnarray}
&&Y^B(\ku,\kd,\kt,\kq)= 
 -\frac{\Cg\Nc\mu^{2 \epsilon }}{\epsilon^2} ((-\sdq)^{-\epsilon}+(-\sut)^{-\epsilon})
 \nonumber \\
 &+&\frac{\Cg\mu^{2 \epsilon}}{\Nc\, \epsilon^2}
  \bigg[(-\sud)^{-\epsilon}+(-\stq)^{-\epsilon}
      +(-\suq)^{-\epsilon}+(-\sdt)^{-\epsilon}
      -(-\sdq)^{-\epsilon}-(-\sut)^{-\epsilon}\bigg]
  \nonumber \\
 &+&\frac{\Cg\mu^{2 \epsilon}}{4 \epsilon} 
   \big(-6 C_f+2 b_0 \big)
        \bigg[(-\sud)^{-\epsilon}+(-\suq)^{-\epsilon}
            +(-\sdt)^{-\epsilon}+(-\stq)^{-\epsilon} \bigg] 
    - \frac{20 n_f}{9}+\frac{80 \Nc}{9}+\frac{8}{\Nc} \nonumber \\
& + &\frac{1}{N_c}\Bigg[\Ls^{2{\rm m}e}_{-1}(\sutq,\sdtq;\stq,\mhsq)
  +  \Ls^{2{\rm m}e}_{-1}(\sudt,\sdtq;\sdt,\mhsq)\nonumber \\
& + &\qquad \Ls^{2{\rm m}e}_{-1}(\sudq,\sutq;\suq,\mhsq)
  +  \Ls^{2{\rm m}e}_{-1}(\sudt,\sudq;\sud,\mhsq)\Bigg] \nonumber \\
& - &(\Nc+\frac{1}{\Nc})\Bigg[\Ls^{2{\rm m}e}_{-1}(\sudt,\sutq;\sut,\mhsq)
  {+}  \Ls^{2{\rm m}e}_{-1}(\sudq,\sdtq;\sdq,\mhsq)\Bigg]\,.
\end{eqnarray}
The finite function $X^B$ is given by 
\begin{eqnarray}
\label{eq:XB}
&& X^B(\ku,\kd,\kt,\kq)= 
      -\Ls_{-1}({{\sud,\sdq;\sudq}}) \; g_{1}(\ku,\kd,\kt,\kq) \;
 \Big(1+\frac{1}{\Nc^2}\Big) 
\nonumber \\
       &-& \Ls_{-1}(\sud,\sdt;\sudt) \; g_{2}(\ku,\kd,\kt,\kq) \;
  \frac{1}{\Nc^2}
\nonumber \\
       &+&  \Ll_1\Big(\frac{-\sudt}{-\sud}\Big) \; g_{3}(\ku,\kd,\kt,\kq) \;
  \Big(1+\frac{1}{\Nc^2}\Big) 
\nonumber \\
       &+&  \Ll_0\Big(\frac{-\sudt}{-\sud}\Big) \; g_{4}(\ku,\kd,\kt,\kq) 
       +  \Ll_0\Big(\frac{-\sudt}{-\sud}\Big) \; g_{5}(\ku,\kd,\kt,\kq) \;
 \frac{1}{\Nc^2}
\nonumber \\
       &+& \ln\Big(\frac{-\sudt}{-\sud}\Big) \; g_{6}(\ku,\kd,\kt,\kq) 
       + \ln\Big(\frac{-\sudt}{-\sud}\Big) \; g_{7}(\ku,\kd,\kt,\kq) \;
    \frac{1}{\Nc^2}
\nonumber \\
       &+& g_{8}(\ku,\kd,\kt,\kq) \; 
    \Big(1+\frac{1}{\Nc^2}\Big)\,, 
\end{eqnarray}
where the functions $g_i$ are given in Appendix~\ref{functionsI}.
We note that the result in Eq.~\eqref{eq:B1} is symmetric under the
exchange of $(1 \leftrightarrow 3)$ or $(2 \leftrightarrow 4)$.

The counterterm renormalizing the ultraviolet divergences in the case of
identical quarks reads
\beq
B_{\rm ct}(\ku,\kd,\kt,\kq)  = 
 -2 \frac{\Cg}{\epsilon} b_0 \frac{g^2}{16 \pi^2} B_0(\ku,\kd,\kt,\kq)
\,,
\eeq
while finite contribution coming from the effective Lagrangian is 
\begin{equation}
B_{\rm fin} (\ku,\kd,\kt,\kq) =  {2} \Delta \; B_0(\ku,\kd,\kt,\kq) 
\,. 
\end{equation}

\subsection{$H \rightarrow q\bar{q}gg$}
At one loop the full amplitude is calculated from 191 Feynman diagrams
which can be decomposed into the 
three color-ordered sub-amplitudes,
\begin{equation}
M_1^C =    (T^{a_3} T^{a_4})_{i_1 i_2} c_1^{(1)}(1,2,3,4)
         +(T^{a_4} T^{a_3})_{i_1 i_2} c_2^{(1)}(1,2,3,4)
         +\delta^{a_3 a_4} \delta_{i_1 i_2} c_3^{(1)}(1,2,3,4)\,.
\end{equation}
Bose symmetry requires that $c_2^{(1)}(1,2,3,4)= c_1^{(1)}(1,2,4,3)$.

The divergent parts of these one-loop amplitudes are given by 
\begin{eqnarray}
c_1^{(1)}(1,2,3,4) \rightarrow
 &&c_\Gamma  \frac{g^2\mu^{2 \epsilon}}{16 \pi^2} 
 \Big[-\frac{N_c}{\epsilon^2}\Big( (-s_{24})^{-\epsilon}+(-s_{13})^{-\epsilon}+(-s_{34})^{-\epsilon}\Big)
 +\frac{1}{N_c \epsilon^2}(-s_{12})^{-\epsilon} \nonumber \\ 
 &-&\frac{3 C_f}{\epsilon}+\frac{b_0}{\epsilon} \Big] c_{1}^{(0)}(1,2,3,4) \\
c_{3}^{(1)}(1,2,3,4) \rightarrow
 && c_\Gamma \frac{g^2 \mu^{2 \epsilon}}{16 \pi^2}
 \Big[\frac{1}{2 \epsilon^2} c_{1}^{(0)}(1,2,3,4)  
 \Big( (-s_{14})^{-\epsilon}+(-s_{23})^{-\epsilon}
       -(-s_{12})^{-\epsilon}-(-s_{34})^{-\epsilon}\Big) 
 \nonumber \\
 &+&\frac{1}{2 \epsilon^2} c_{2}^{(0)}(1,2,3,4) 
\Big( (-s_{13})^{-\epsilon} +(-s_{24})^{-\epsilon} 
     -(-s_{12})^{-\epsilon} -(-s_{34})^{-\epsilon} \Big)\Big]\,.
\end{eqnarray}

The interference between the Born and the NLO amplitude is
given by
\begin{eqnarray}
2 \; \mbox{Re}(M_1^C M_0^{C \; \star}) &=& 
\frac{V N_c}{2} \mbox{Re} [c_1^{(1)} c_{1}^{(0)\;\star}+c_2^{(1)} c_{2}^{(0)\;\star}]
\nonumber \\
&-&\frac{V}{2 N_c} \mbox{Re} [(c^{(1)}_1+c^{(1)}_2) (c_{1}^{(0)}+c_{2}^{(0)})^\star]
+V \mbox{Re} [ c^{(1)}_{3} (c_{1}^{(0)}+c_{2}^{(0)})^\star]\,,
\end{eqnarray}
with $c_i \equiv  c_i(1,2,3,4)$.
Counterterms, analogous to those in 
Eqs.~(\ref{counterterm1}, \ref{counterterm2})
need to be included to obtain the full renormalized result.

Numerical results, which are given in the following section, were
generated using an extension of the method suggested in
ref.~\cite{Giele:2004iy}.  Analytic expressions for the Feynman graphs
are generated using Qgraf~\cite{Nogueira:1991ex} and
Form~\cite{Vermaseren:2000nd}. The scalar and tensor integrals
appearing in the amplitudes are reduced numerically using the
Davydychev reduction for the tensor integrals~\cite{Davydychev:1991va}
and a recursive procedure similar to the one proposed
in~\cite{Giele:2004iy} to reduce all scalar integrals to a small
number of analytically known basis integrals.  These are then
evaluated numerically as a Laurent series in the $\epsilon$
parameter\footnote{The {\it numerical} Laurent expansion technique was
  first used in ref.~\cite{vanHameren:2005ed}. In a more general
  analytic context it was used by many authors before.}.  The key
point of this method is that a record is kept of all previously
computed integrals, so that each scalar integral is computed only
once.  The result of our procedure is a numerical expression for the
scalar and tensor integrals component by component each of which has a
Laurent expansion in $\epsilon$.  This method will be described in
detail in a later paper~\cite{egz2}.  Numerical or semi-numerical
methods have also been described in
refs.~\cite{vanHameren:2005ed,Binoth:2005ff,Binoth:2002xh,Ferroglia:2002mz,Andonov:2004hi,Belanger:2003sd,deDoncker:2004bf,Nagy:2003qn,delAguila:2004nf}.

\subsection{$H \rightarrow gggg$}

At NLO the amplitude for process Eq.~(\ref{Hgggg}) 
requires the calculation of 739 Feynman diagrams, which 
can expanded in nine color sub-amplitudes 
\begin{eqnarray} \label{GeneralExpansion}
M_1^D &=& 
\sum_{\sigma \in S_4/Z_4} 
\mbox{tr}(T^{a_{\sigma(1)}} T^{a_{\sigma(2)}} T^{a_{\sigma(3)}} T^{a_{\sigma(4)}}) \; d_{1}^{(1)}(\sigma(1),\sigma(2),\sigma(3),\sigma(4))\nonumber \\
 &+& \frac{1}{N_c}\mbox{tr}(T^{a_1} T^{a_2})\; \mbox{tr}(T^{a_3} T^{a_4}) \; d_2^{(1)}(1,2,3,4) \nonumber \\
 &+& \frac{1}{N_c}\mbox{tr}(T^{a_1} T^{a_3}) \; \mbox{tr}(T^{a_2} T^{a_4}) \; d_2^{(1)}(1,3,2,4) \nonumber \\
 &+& \frac{1}{N_c} \mbox{tr}(T^{a_1} T^{a_4}) \; \mbox{tr}(T^{a_2} T^{a_3}) \; d_2^{(1)}(1,4,2,3)\,.
\end{eqnarray}

If we discard diagrams with internal quark loops we have the 
decoupling identity~\cite{Bern:1990ux}
\beq \label{decoupling}
d_2^{(1)}(1,2,3,4)=  \sum_{\sigma \in S_4/Z_4} 
 d_1^{(1)}(\sigma(1),\sigma(2),\sigma(3),\sigma(4))\,.
\eeq
However, at NLO the $d_2$ terms in Eq.~(\ref{GeneralExpansion})
do not receive contributions from internal fermion loops. This can be easily
shown by explicitly examining the diagrams with internal 
fermionic bubbles, triangles, and boxes. 
The general expansion can thus be simplified as a consequence of 
Eq.~(\ref{decoupling}) so that
\begin{eqnarray} \label{ParticularExpansion}
M_1^D &=& 
\sum_{\sigma \in S_4/Z_4} 
\mbox{tr}(T^{a_{\sigma(1)}} T^{a_{\sigma(2)}} T^{a_{\sigma(3)}} T^{a_{\sigma(4)}}) d_{1}^{(1)}(\sigma(1),\sigma(2),\sigma(3),\sigma(4)) \nonumber \\
 &+& \frac{1}{N_c}\Big[
  \mbox{tr}(T^{a_1} T^{a_2}) \; \mbox{tr}(T^{a_3} T^{a_4})
 +\mbox{tr}(T^{a_1} T^{a_3}) \; \mbox{tr}(T^{a_2} T^{a_4})
\nonumber \\
 \qquad &+& \mbox{tr}(T^{a_1} T^{a_4}) \; \mbox{tr}(T^{a_2} T^{a_3}) \Big] \; 
d_2^{(1)}(1,2,3,4) \; . 
\end{eqnarray}
Using Eq.~(\ref{ParticularExpansion})
it can be shown that the result 
for the matrix element squared is
\begin{eqnarray}
&&|M_0^D+M_1^D|^2-|M_1^D|^2 = 
\frac{N_c^2 (N_c^2-1)}{16} \sum_{\sigma \in S_4/Z_4} 
\Big\{ |d_{1}^{(0)}(\sigma(1),\sigma(2),\sigma(3),\sigma(4))|^2\nonumber \\
 &+&2 \; \mbox{Re} \Big[d_{1}^{(0)}(\sigma(1),\sigma(2),\sigma(3),\sigma(4))^{\star}
 \; d_1^{(1)}(\sigma(1),\sigma(2),\sigma(3),\sigma(4)) \Big] \Big\}\,.
\end{eqnarray}
Counterterms, analogous to those in 
Eqs.~(\ref{counterterm1}, \ref{counterterm2})
need to be included to obtain the full renormalized result.

Numerical results for this matrix element squared were generated 
using the method described above.
The pole structure for the color sub-amplitude $d_1$ 
has the simple form
\beq
d_1^{(1)}(1,2,3,4) \rightarrow  
\frac{c_\Gamma \; g^2 \mu^{2 \epsilon} }{16 \pi^2} 
\Big[ -\frac{N_c}{\epsilon^2} \Big( 
  (-s_{12})^{-\epsilon} +(-s_{23})^{-\epsilon} 
 +(-s_{34})^{-\epsilon} +(-s_{14})^{-\epsilon} \Big) \Big] 
d_{1}^{(0)}(1,2,3,4) \; .
\eeq

\section{Numerical results} 
\label{numer}
In this section we present numerical results for the Born amplitude squared
and for its interference with the 
one-loop matrix element for the four processes of interest,
$A,B,C$ and $D$. We use the following arbitrarily chosen, momentum configuration, where a Higgs boson 
of unit mass decays into four well separated partons, $(E,p_x,p_y,p_z)$:
\begin{equation}
\label{momentumchoice}
\begin{array}{lllll}
p_H=& (  -1.00000000000,& \phantom{+}0.00000000000, & \phantom{+}0.00000000000,& \phantom{+}0.00000000000)\,, \\ 
k_1=& ( + 0.30674037867,& -0.17738694693,& -0.01664472021,& -0.24969277974)\,, \\ 
k_2=& ( + 0.34445032281,& +0.14635282800, & -0.10707762397,& +0.29285022975)\,, \\ 
k_3=& ( + 0.22091667641,& +0.08911915938, & +0.19733901856,& +0.04380941793)\,, \\ 
k_4=& ( + 0.12789262211,& -0.05808504045,& -0.07361667438,& -0.08696686795)\,.
\end{array}
\end{equation}
For each process, \{$A, B, C, D$\}, we introduce the quantities
\begin{eqnarray}
X_{B} &=& \frac{1}{g^4 A^2} X_0(\ku,\kd,\kt,\kq)\,,\nonumber\\
X_{V} &=& \frac{8\pi^2}{ g^6 A^2}\Big[X_1(\ku,\kd,\kt,\kq)-X_0(\ku,\kd,\kt,\kq)\Big]\,,
\qquad \mbox{with } X=A,B,C,D\,, 
\end{eqnarray}
which are independent of the value of the coupling constant.
Thus $X_B$ is the 
matrix element squared evaluated using the Born amplitude.
$X_{V,N}$ and $X_{V,A}$ denote the contributions of 
the interference between the virtual amplitude and the lowest order,
as calculated from the numerical and
analytical formulas.
The unrenormalized results are given in Table~\ref{Numerical}
for the scale choice $\mu=M_H$ and the momenta
of Eq.~(\ref{momentumchoice}).
\begin{table}

\begin{center}
\begin{tabular}{|c|c|c|c|}
\hline 
        &  $\Cg/\epsilon^2$ &$\Cg/\epsilon $ &1 \\
\hline \hline 
$A_B$   & 0 & 0 & 12.9162958212387\\ 
%CDR scheme 
%$A_{V,N}$& -68.8869110466064 & 7.50924350098691 & 323.304110584114\\
%$A_{V,A}$& -68.8869110466063 & 7.50924350099082 & 323.304110584136\\
%HV scheme 
$A_{V,N}$&-68.8869110466064&-114.642248172523&120.018444115429\\
$A_{V,A}$&-68.8869110466063& -114.642248172519& 120.018444115458\\
\hline \hline
% CDR scheme 
%$B_B$   & 0 & 0 & 858.856417157052\\ 
%$B_{V,N}$& -4580.56755817099 & 8663.59943211762 & 27919.5795296170\\
%$B_{V,A}$& -4580.56755817094 & 8663.59943211806 & 27919.5795296170\\
%HV scheme 
$B_B$   & 0 & 0 & 858.856417157052\\ 
$B_{V,N}$&-4580.56755817099  &-436.142317955660  &26470.9608978346\\
$B_{V,A}$&-4580.56755817094  &-436.142317955208   &26470.9608978350\\
\hline \hline
$C_B$   & 0 & 0 &968.590160211857 \\ 
$C_{V,N}$& -8394.44805516930 &-19808.0396331354  & -1287.90574949112\\
$C_{V,A}$& -8394.44805516942 &-19808.0396331363  & \mbox{not known} \\
\hline \hline
$D_B$   & 0 & 0 &3576991.27960852 \\ 
$D_{V,N}$&   -4.29238953553022    $\cdot 10^7$ &-1.04436372655580 $\cdot 10^8$ & -6.79830911471604$\cdot 10^7$ \\
$D_{V,A}$&   -4.29238953553022$\cdot 10^7$ & -1.04436372655580 $\cdot 10^8$ & \mbox{not known}  \\
\hline
\end{tabular}
\end{center}
\caption{\label{Numerical}
Numerical results for the Born amplitude squared, ($X_B$), 
and the numerical and analytic one-loop corrections, ($X_{V,N}$ and $X_{V,A}$),
to the four processes $A,B,C,D$, Eqs.~(\ref{fourprocesses1}--\ref{fourprocesses4}).}
\end{table}

The explicit results show that far from exceptional momentum
configurations, where divergent inverse Gram determinants are known to
spoil the accuracy of the numerical procedure, a relative accuracy of
$\cO{10^{-13}}$ can be achieved.  For processes $C$ and $D$, where a
full analytical result is not available, we verified that the answer
satisfies the Ward identities to a similar relative accuracy.  For
process $D$ we checked numerically that for $n_f=0$, the color
amplitudes satisfy the decoupling identity, Eq.~\eqref{decoupling}.
Close to exceptional momentum configurations, it is still possible to
use a numerical approach\cite{Giele:2004ub,egz2}.

We have also checked numerically that our results satisfy the following 
relationship between the HV and FDH regularization schemes,
\begin{eqnarray}
a_1^{(1)\;\mbox{\scriptsize FDH}}(1,2,3,4)-a_1^{(1)\;\mbox{\scriptsize HV}}(1,2,3,4)&=& 
\frac{g^2}{16 \pi^2} \Big(\frac{\Nc}{3}-\frac{1}{\Nc}\Big) a^{(0)}(1,2,3,4)   \; ,
\nonumber \\
a_2^{(1)\;\mbox{\scriptsize FDH}}(1,2,3,4)-a_2^{(1)\;\mbox{\scriptsize HV}}(1,2,3,4)&=& 0  \; ,
\nonumber \\
c_1^{(1)\;\mbox{\scriptsize FDH}}(1,2,3,4)-c_1^{(1)\;\mbox{\scriptsize HV}}(1,2,3,4)&=& 
\frac{g^2}{16 \pi^2} \Big(\frac{\Nc}{6}-\frac{1}{2 \Nc}\Big) c^{(0)}_1(1,2,3,4)  \; , 
\nonumber \\
c_3^{(1)\;\mbox{\scriptsize FDH}}(1,2,3,4)-c_3^{(1)\;\mbox{\scriptsize HV}}(1,2,3,4)&=& 0 \; ,
\nonumber \\
d_1^{(1)\;\mbox{\scriptsize FDH}}(1,2,3,4)-d_1^{(1)\;\mbox{\scriptsize HV}}(1,2,3,4)&=& 0 \; .
\label{UV+IR}
\end{eqnarray}
Applying the finite renormalization which compensates for the difference between the ultraviolet
regularization in the two schemes \cite{Kunszt:1993sd}, we recover the expected 
difference between the two schemes due to the differing infrared regularization, 
\begin{eqnarray}
a_1^{(1)\;\mbox{\scriptsize FDH}}(1,2,3,4)-a_1^{(1)\;\mbox{\scriptsize HV}}(1,2,3,4)&=& 
\frac{g^2}{4 \pi^2} \tilde{\gamma}_q\,  a^{(0)}(1,2,3,4)   \; ,
\nonumber \\
c_1^{(1)\;\mbox{\scriptsize FDH}}(1,2,3,4)-c_1^{(1)\;\mbox{\scriptsize HV}}(1,2,3,4)&=& 
\frac{g^2}{8 \pi^2} (\tilde\gamma_q +\tilde\gamma_g)\, c^{(0)}_1(1,2,3,4)  \; , 
\nonumber \\
d_1^{(1)\;\mbox{\scriptsize FDH}}(1,2,3,4)-d_1^{(1)\;\mbox{\scriptsize HV}}(1,2,3,4)&=& 
\frac{g^2}{4 \pi^2} \tilde\gamma_g\, d^{(0)}_1(1,2,3,4)  \; ,
\end{eqnarray}
where
\beq
\tilde\gamma_q \equiv  \frac{C_f}{2}\, \quad \mbox{and} \quad 
\tilde \gamma_g \equiv \frac{N_c}{6}\,. 
\eeq
The other two relations in Eq.~(\ref{UV+IR}) are unchanged.

\section{Outlook}

In this paper we presented results obtained using a general,
semi-numerical calculation of one-loop corrections. In order to
establish the feasibility of the semi-numerical method, we computed
all the one-loop corrections to Higgs plus four parton processes using
an effective Lagrangian.  We presented explicit results for a
specific, non-exceptional phase space point. For practical
applications of this method, one has to be able to treat exceptional
momentum configurations also. The method of this paper can be extended
to the treat those regions. A detailed description of the algorithm is
presented in a separate work~\cite{egz2}.

The results presented in this paper generate two separate lines of
research. The first is clearly the completion of the calculation of
the Higgs boson plus two jet process at next-to-leading order. As
indicated in the text all of the needed elements are now in place.

The second development is the extension of these methods to calculate
other one-loop processes which currently lie beyond the range of
analytic calculation. Examples of processes of current experimental
interest are diboson plus one jet ($V_1,V_2,j$), tri-boson production
($V_1,V_2,V_3$) and vector boson plus heavy quark pairs ($V Q
\bar{Q}$).

\subsection*{Acknowledgments} 

We are happy to acknowledge useful discussions with W.A.~Bardeen,
E.W.N.~Glover and U.~Haisch. We thank Carola Berger and Lance Dixon
for pointing out typos in the analytical expressions of four quark
amplitudes, which have been corrected in the present version of the
paper.

\appendix

\section{Integral Functions Appearing in Amplitudes}
\label{functionsA}
The integral functions appearing in the virtual corrections 
are presented in this appendix. Following closely the notation 
of ref.~\cite{Bern:1997sc} we define
\beq
\Ll_0(r) = \frac{\ln(r)}{1-r}\,,  \qquad 
\Ll_1(r) = \frac{\Ll_0(r)+1}{1-r} \; .
\eeq
The above functions have the property that they are finite as 
their denominators vanish.  
Furthermore we define 
\begin{eqnarray}
\Ls_{-1}(s,t;m^2) &=&
      \li(1-\frac{s}{m^2}) + \li(1-\frac{t}{m^2}) 
 + \ln \frac{-s}{-m^2}\,\ln \frac{-t}{-m^2}
  - \frac{\pi^2}{6}\,,
\end{eqnarray}
where the dilogarithm is defined as usual as 
\beq
\li(x) = - \int_0^x dz \, \frac{\ln(1-z)}{z}\,.
\eeq
The function $\Ls_{-1}$ is simply related to the scalar box integral 
with one external mass evaluated in six space-time dimensions, where 
it is infrared- and ultraviolet-finite.

The `easy' six-dimensional box function with two non-adjacent external
masses, $m_1,m_3$, is related to the function $\Ls^{2{\rm m}e}_{-1}$
\begin{eqnarray}
  \Ls^{2{\rm m}e}_{-1}(s,t;m_1^2,m_3^2) &= 
    -\li\left(1-\frac{m_1^2}{ s}\right)                      
    -\li\left(1-\frac{m_1^2}{ t}\right)                      
    -\li\left(1-\frac{m_3^2}{ s}\right)                      
    -\li\left(1-\frac{m_3^2}{ t}\right) \cr                     
&
    +\li\left(1-\frac{m_1^2 m_3^2 }{ st}\right)
    -\frac{1}{2}\ln^2\left(\frac{-s}{-t}\right) \; .
\end{eqnarray}
This function has the property that it vanishes as $s+t-m_1^2-m_3^2
\to 0$. The analytic continuation of these integrals is obtained 
adding a small positive imaginary part to each
invariant, $s_{ij} \to s_{ij}+i \varepsilon $.

\section{Functions for distinct quarks}
\label{functionsD}
The kinematic functions for the virtual corrections to $H \rightarrow
q \bar{q} q' \bar{q}'$ appearing in Eq.~\eqref{eq:XA} are given below:
\begin{eqnarray}
f_{1}(\ku,\kd,\kt,\kq) &=& 
  -\frac{\sud \stq}{2 \sut^2}
 +\frac{3 \sut \sdq-\sdt^2+\suq \sdt-\suq^2-\sut^2}{\sud \stq} 
 - \frac{\suq^2 \sdt^2}{2 \sud \sut^2 \stq} \nonumber \\
 &-&\frac{\sdq^2}{2 \sud \stq}
 -2 \frac{(\sut \sdq-\suq \sdt)^2}{\sud^2 \stq^2}
    +\frac{\sdq}{\sut}+\frac{\suq \sdt}{\sut^2}-2 \\
f_{2}(\ku,\kd,\kt,\kq) &=& 
   \frac{\sud \stq (\sud \stq+\sdt (\sdq+2 \sdt-\suq))+\sdt^2 (\sdq+\suq)^2}
      {2 \sud^3 \stq}
\\ 
f_{3}(\ku,\kd,\kt,\kq) &=& 
   \frac{\stq}{2 \sdt}+\sdt (\sdq+\suq) \frac{\sdq+4 \sdt+3 \suq}
{2 \sud^2 \stq} +\frac{3 \sdq+4 \sdt}{2 \sud}
\\
f_{4}(\ku,\kd,\kt,\kq) &=&  
  -2 \frac{\stq}{\sdt}-\sdt (\sdq+\suq) \frac{\sdq+2 \sdt+5 \suq}
 {2 \sud^2 \stq}
\nonumber \\
                &-&\frac{4 \sdq+6 \sdt-3 \suq}{2 \sud}
\\
f_{5}(\ku,\kd,\kt,\kq) &=&
-\frac{\sut}{\suq}  - \frac{2 \sdt}{\sdq} + \frac{\sdq}{\suq} - \
\frac{\sdt}{\stq} - \frac{\suq \sdt^2}{\sdq^2 \stq} + \frac{\sut \sdt}{\sdq \stq} + \
\frac{\sut \sdq}{\suq \stq} - \frac{\stq}{\suq}
\\
f_{6}(\ku,\kd,\kt,\kq)&=&
  \frac{\sud \stq}{2 \sut \sdt}+
 \frac{4 \sdt \sdq+2 \suq \sdq-3 \sut \sdq+3 \suq \sdt}
                          {2 \sud \stq} \nonumber \\
&+&\frac{\suq^2 \sdt}{2 \sud \sut \stq}-\frac{\suq}{\sut}+\frac{1}{2}
\\
f_{7}(\ku,\kd,\kt,\kq)&=&
        \frac{\suq^2 \sdt^2+\sut \stq^2 \sud-\sut \suq \sdt \sdq}
 {2 \sut^2 \stq \sud}
\\
f_{8}(\ku,\kd,\kt,\kq)&=&
         \frac{\suq \sdt-\sut \sdq}{2 \sud \stq}\\
f_{9}(\ku,\kd,\kt,\kq)&=& 
    \frac{\sut \sdq^2}{\sud \sdt \stq}
   -\frac{\sud \stq}{\sut \sdt}
  -\frac{\suq^2 \sdt}{\sud \sut \stq}-2 \frac{\sdq}{\sdt}
 +2 \frac{\suq}{\sut}-1
\nonumber \\
 &-&\frac{2 \sdq^2+2 \sdt \sdq+5 \suq \sdq-5 \sut \sdq+5 \suq \sdt}
                         {2 \sud \stq} \\
\\
f_{10}(\ku,\kd,\kt,\kq)&=& 
     \frac{\sud \sdt \stq^2+\sut^2 \sdq^2-\sut \suq \sdt \sdq} 
 {\sud \sdt^2 \stq}
\\
f_{11}(\ku,\kd,\kt,\kq)&=& \frac{2 \sut \sdq}{\sud \stq} 
\\
f_{12}(\ku,\kd,\kt,\kq)& = & 
   \frac{\sut (\sut (\suq-\sdq)+2 \suq \sdt)}{ 2 \sud^2 \stq}
 +\frac{\suq}{2 \sud} \; .
\end{eqnarray}

\section{Functions for identical quarks}
\label{functionsI}
The kinematic functions for the virtual corrections to $H \rightarrow
q \bar{q} q \bar{q}$ appearing in Eq.~\eqref{eq:B1} are given below:
\begin{eqnarray}
g_{1}(\ku,\kd,\kt,\kq) &=& -1 - \frac{\sut \sdq\left( \sut^2 + \sdq^2 \right) }{
4 \sud \suq \sdt \stq} + \frac{\sut^2 + 2 \suq \sdt - 2 \sut \sdq + \sdq^2}{4 \sud \stq} \nonumber \\ &+& 
\frac{\sut^2 - 2 \sut \sdq + \sdq^2 + 2 \sud \stq}{4 \suq \sdt}\\
g_{2}(\ku,\kd,\kt,\kq)
 &=& \frac{1}{4} - \frac{\sdq}{4 \sut} - \frac{\suq^2 \sdt
^2}{8\sud \sut^2 \stq} + \frac{3\, \suq \sdt \sdq}{8\sud \sut \stq} + \frac{\sut \sdq\left( \sut^
2 + \sdq^2 \right) }{4 \sud \suq \sdt \stq} \nonumber \\ 
&-& \frac{2 \sut^2 + 3\, \suq \sdt - 3\, \sut \sdq + 4
\sdq^2}{8\,\sud \stq} + \frac{3\, \sud \sdq \stq}{8\sut \suq \sdt} 
- \frac{\sud^2 \stq^2}{8\sut^2 \suq \sdt} \nonumber \\
&+& \frac{\suq \sdt + \sud \stq}{8\sut^2} 
- \frac{2 \sut^2 - 3\, \sut \sdq + 4 \sdq^2 +3\, \sud \stq}{8\suq \sdt}\\
g_{3}(\ku,\kd,\kt,\kq)&=& \frac{\sdt^2 \sdq}{8\sud^2 \suq} + \frac{\sdt\left( \suq + \sdt + \sdq \right) }{8\sud^2} - \frac{\stq}{4 \sud} + \frac{\left( \sdt + \sdq 
\right) \stq}{8\sud \suq} + \frac{\stq^2}{8\suq \sdt}\\
g_{4}(\ku,\kd,\kt,\kq) &=& \frac{6\, \sdt - 3\, \sdq}{8\sud} + \frac{\sdq\left( -\sdt
+ 4 \sdq \right) }{8\sud \suq} - \frac{\stq}{4 \suq} + \frac{5\, \sdq \stq}{8\, \suq \sdt}\\
g_{5}(\ku,\kd,\kt,\kq) &=& \frac{\suq}{4 \sud} + \frac{\sdq}{8\sud} - \frac{\sdq^
2}{4 \sud \suq} + \frac{3\, \sdt\left( -2 \suq + \sdq \right) }{8\sud \suq} - \frac{\sdt{
\left( \suq + \sdq \right) }^2}{4 \sud^2 \stq} \nonumber \\
&+& \frac{\stq}{4 \suq} + \frac{\stq}{4 \sdt} 
- 
\frac{3\, \sdq \stq}{8\suq \sdt} - \frac{\sud \stq^2}{4 \suq \sdt^2}\\
g_{6}(\ku,\kd,\kt,\kq) &=& \frac{5}{8} - \frac{\sdq}{4 \suq} + \frac{3\, \sut \sdq
 + 4 \sdq^2 - 3\, \sud \stq}{8\suq \sdt}\\
g_{7}(\ku,\kd,\kt,\kq) &=& - \frac{7}{8}  + \frac{\suq}{4 \sut}
+ \frac{\sdq}{4 \suq} - \frac{\suq^2 \sdt}{4 \sud \sut \stq} - \frac{\sdq^2}{4 \sud \stq} + 
\frac{\sud \stq}{4 \sut \sdt} - \frac{\sud^2 \stq^2}{4 \sut \suq \sdt^2} \nonumber \\
&-& \frac{\sut \sdq + 2
\sdq^2 - \sud \stq}{8\suq \sdt}\\
g_{8}(\ku,\kd,\kt,\kq) &=& 
\frac{\sud}{32 \suq} + \frac{\suq}{32 \sud} + \frac{\sud}
{32 \sdt} + \frac{\sut\left( \suq - 2 \sdq \right) }{64 \sud \sdt} - \frac{\sut \sdq}{32
 \sud \suq} \nonumber \\   
&+& \frac{\sut \sdq\left( \sut + \sdq \right) }{64 \sud \suq \sdt} 
 + \frac{\sdt
\left( 2 \suq + \sdq \right) }{64 \sud \suq} 
+ \frac{\suq + \sdt}{32 \stq} + \frac{\sut
\left( \sud + \sdt - 2 \sdq \right) }{64 \suq \stq} 
\nonumber \\
&+& \frac{\left( \sud - 2 \sut + \suq \right) \sdq
}{64 \sdt \stq} 
+ \frac{\sut \sdq\left( \sut + \sdq \right) }{64 \sud \suq \stq} 
+ \frac{\sut
 \sdq\left( \sut + \sdq \right) }{64 \sud \sdt \stq} \nonumber \\ 
&+& \frac{\sut \sdq\left( \sut + \sdq 
\right) }{64 \suq \sdt \stq} + \frac{\left( \sut \suq + 2 \sud\left( \suq + \sdt \right)  + \sdt
 \sdq \right) \stq}{64 \sud \suq \sdt} \; .
\end{eqnarray}

%%%%%%%%%%%%%%%%%%%%%%%%%%%%%%%%%%%%%%%%%%%%%%%%%%%%%%

\end{document}